\documentclass[11pt,letterpaper]{article}
\usepackage[margin=1in]{geometry}
\usepackage{amsmath,amssymb,amsthm}
\usepackage{graphicx,booktabs,placeins}
\usepackage{tikz}
\usetikzlibrary{arrows.meta}
\usepackage[round,authoryear]{natbib}
\usepackage{xurl}
\usepackage{microtype}
\usepackage[hidelinks]{hyperref}
\hypersetup{pdftitle={Aalen--Johansen as an augmented IPCW estimator},pdfauthor={Lu Mao}}
\newtheorem{theorem}{Theorem}
\newtheorem{proposition}{Proposition}
\theoremstyle{definition}
\newtheorem{definition}{Definition}
\newlength{\figwidth}
\newenvironment{tabnote}{\par\vskip7pt\footnotesize}{\par}
\newcommand{\pr}{\operatorname{pr}}
\newcommand{\prob}{\operatorname{pr}}
\newcommand{\E}{E}
\newcommand{\var}{\operatorname{var}}
\newcommand{\cov}{\operatorname{cov}}
\newcommand{\dd}{\,\mathrm{d}}
\newcommand{\ipcw}{\text{\upshape\textsc{ipcw}}}

\title{Aalen--Johansen as an augmented IPCW estimator}
\author{Lu Mao\\[0.35em]
  \small Department of Biostatistics and Medical Informatics\\
  \small University of Wisconsin--Madison\\
  \small 610 Walnut Street, Madison, Wisconsin 53726, U.S.A.\\
  \small \href{mailto:lmao@biostat.wisc.edu}{\texttt{lmao@biostat.wisc.edu}}}
\date{}

\begin{document}
\maketitle

\begin{abstract}
We show that, in every sample, the widely used Aalen--Johansen estimator of a state
occupation probability is an inverse probability of censoring weighted
estimator augmented by a working Markov prediction from each subject's
current state. This identity explains the robustness of the Aalen--Johansen
against departures from the Markov assumption. It also reveals that its influence
function differs from the semiparametric efficient one by two orthogonal
components: the information in earlier history beyond the current state, and
the error of the Markov prediction. Based on this decomposition, we propose a
practical, model-free least-squares augmentation of the Aalen--Johansen that regresses the outcome on a history score, narrowing the gap under non-Markov processes without increasing the asymptotic variance. Its advantages are validated both theoretically and in simulations.
\end{abstract}


\section{Background}\label{sec:motivation}
\begin{figure}[!h]
\centering
\setlength{\figwidth}{0.97\linewidth}
\begin{tikzpicture}[
  x=1pt,y=1pt,>=latex,line cap=round,line join=round,
  state/.style={draw=black,rounded corners=1.6pt,line width=.4pt,
    text=black,font=\small,align=center,inner xsep=5pt,inner ysep=3pt,
    minimum height=17pt},
  panel/.style={text=black,font=\small\itshape,anchor=west,inner sep=0pt},
  transition/.style={->,line width=.6pt,shorten <=1pt,shorten >=1pt}]
  \node[panel] at (0,0) {(a) Life--death};
  \node[state] (alive) at (38,-35) {0. Alive};
  \node[state] (dead)  at (151,-35) {1. Dead};
  \draw[transition] (alive) -- (dead);

  \node[panel] at (207,0) {(b) Competing risks};
  \node[state] (atrisk) at (248,-35) {0. Alive};
  \node[state] (cause1) at (358,-20) {1. Cancer death};
  \node[state] (cause2) at (358,-50) {2. Other-cause death};
  \draw[transition] (atrisk) -- (cause1);
  \draw[transition] (atrisk) -- (cause2);

  \node[panel] at (0,-75) {(c) Illness--death};
  \node[state] (remission) at (38,-100) {0. Remission};
  \node[state] (relapse)   at (151,-100) {1. Relapse};
  \node[state] (death)     at (94.5,-146) {2. Dead};
  \draw[transition] (remission) -- (relapse);
  \draw[transition] (remission) -- (death);
  \draw[transition] (relapse) -- (death);

  \node[panel] at (207,-75) {(d) Liver disease};
  \node[state] (compensated)   at (248,-100) {0. Compensated\\cirrhosis};
  \node[state] (decompensated) at (358,-100) {1. Decompensated\\cirrhosis};
  \node[state] (transplant)    at (248,-146) {2. Alive after\\transplantation};
  \node[state] (liverdeath)    at (358,-146) {3. Dead};
  \draw[transition] (compensated) -- (decompensated);
  \draw[transition] (compensated) -- (transplant);
  \draw[transition] (decompensated) -- (transplant);
  \draw[transition] (decompensated) -- (liverdeath);
  \draw[transition] (transplant) -- (liverdeath);
\end{tikzpicture}
\caption{Examples of multistate models.}
\label{fig:models}
\end{figure}
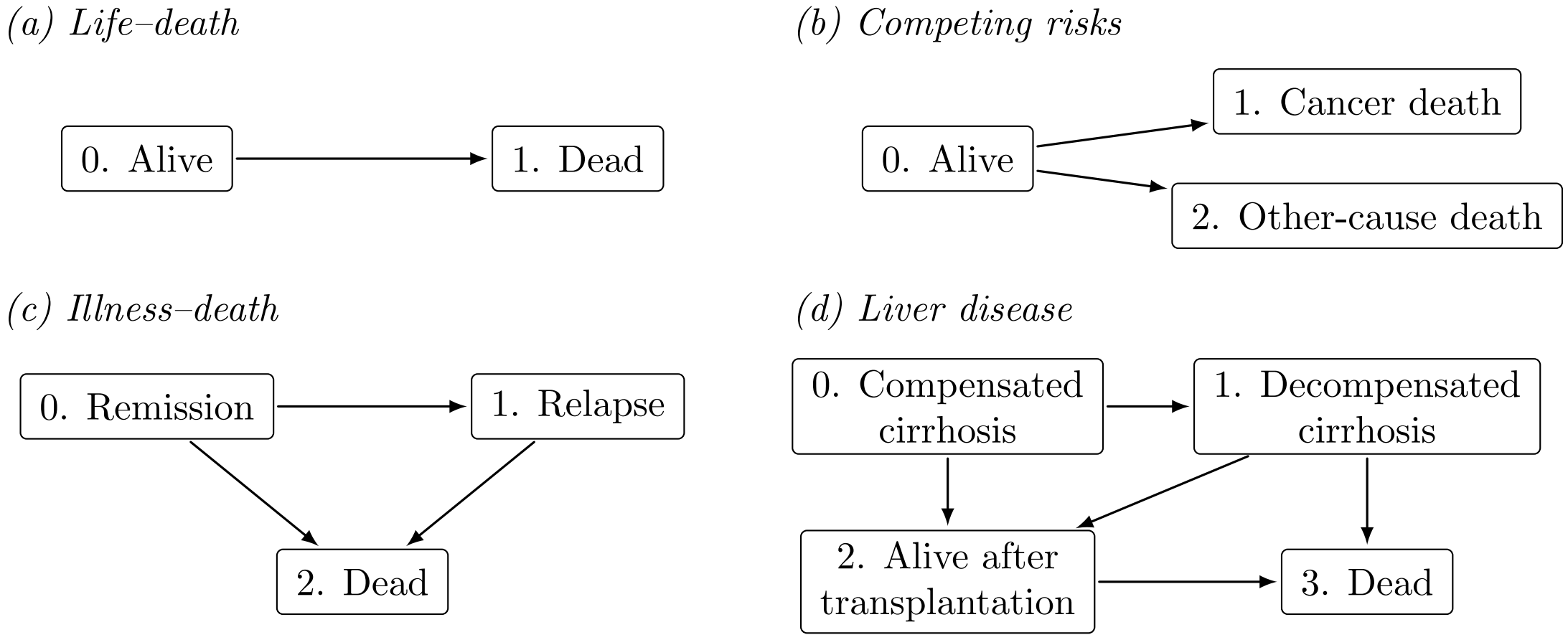

For a multistate process $Y$ with right-continuous sample paths, let $Y(t)\in\{0,\ldots,K\}$ denote a subject's state at time $t$. Figure~\ref{fig:models} shows common examples. Let $p(t)=\{p_0(t),\ldots,p_K(t)\}^{\mathsf T}$, with $p_k(t)=\prob\{Y(t)=k\}$, denote the column vector of state occupation probabilities.

Under right censoring, \citet{aalen1978} estimated $p(t)$ nonparametrically under a Markov model, in which the instantaneous transition probability depends on the history only through the current state:
\begin{equation}\label{eq:markov}
\begin{split}
\text{Markov assumption:}\quad
 \prob\{Y(t+\mathrm d t) = k\mid Y(t-)=j, \mathcal H(t-)\}&=\\
 \prob\{Y(t+\mathrm d t) = k\mid Y(t-)=j\}&\equiv\mathrm d\Lambda_{jk}(t),
\end{split}
\end{equation}
where $\mathcal H(t)=\{Y(s):0\le s\le t\}$ and $\mathrm d\Lambda_{jk}(t)$ is the $j\to k$ transition intensity, collected in the matrix $\mathrm d\Lambda(t)$ with $\mathrm d\Lambda_{jj}(t)=-\sum_{k\ne j}\mathrm d\Lambda_{jk}(t)$. Under \eqref{eq:markov}, the transition probability matrix $P(s,t)$, with $(j,k)$ entry $P_{jk}(s,t)=\prob\{Y(t)=k\mid Y(s)=j\}$, is expressed as the product integral
\begin{equation}\label{eq:prodint}
 P(s,t)=\prod_{s<u\le t}\{I+\mathrm d\Lambda(u)\}
\end{equation}
\citep[Section~2.2.1]{cook2018}. Simple algebraic manipulation yields $p(t)=P(0,t)^{\mathsf T}p(0)$.

With censored data, we can estimate $\mathrm d\Lambda_{jk}(t)$ by the 
Nelson--Aalen-type increment $\mathrm d\widehat\Lambda_{jk}(t)$,
defined as the observed $j\to k$ transitions at $t$ over the number under observation in $j$ just before $t$, 
and $p(0)$ by the empirical proportions $\widehat p(0)$.
This leads to the Aalen--Johansen estimator as a plug-in estimator:
\begin{equation}\label{eq:product}
\text{Aalen--Johansen:}\quad
 \widehat p(t)=\widehat P(0,t)^{\mathsf T}\widehat p(0),\quad\text{where}\;
 \widehat P(s,t)=\prod_{s<u\le t}\{I+\mathrm d\widehat\Lambda(u)\}.
\end{equation}
Although \eqref{eq:product} is derived from a Markov model, $\widehat p(t)$ stays consistent under independent censoring even when the Markov assumption
\eqref{eq:markov} fails to hold \citep{datta2001,glidden2002}. This robustness and its simplicity have made it the standard nonparametric estimator of state occupation probabilities \citep{putter2007} and their derived quantities such as occupation times \citep{buhler2023}.

In fact, the Aalen--Johansen reduces to familiar estimators in simple cases. 
In the life--death model of Figure~\ref{fig:models}(a), $\widehat p_0(t)$ is equivalent to the Kaplan--Meier survival estimator. In the competing risks model of Figure~\ref{fig:models}(b), $\widehat p_k(t)$ is the usual nonparametric estimator of the cumulative incidence of cause $k$ \citep{andersen1993}. 
Both can be reformulated as inverse probability of censoring weighted (IPCW) estimators, which estimate $p_k(t)$ cross-sectionally by averaging $I\{Y(t)=k\}$ over the subjects whose state at $t$ is known, each weighted by the inverse of its probability of remaining uncensored, estimated by the Kaplan--Meier estimator of the censoring time \citep{satten2001,zhang2009}.

This equivalence does not carry over to more complex settings. After all, 
the IPCW estimator uses only whether a subject occupies $k$ at $t$, whereas \eqref{eq:product} is cumulative, propagating the transition rates from 0 to $t$ and so using the states visited along the way. In Figure~\ref{fig:models}(c), for instance, 
the IPCW estimator of the probability of being dead at $t$ considers
only the current state and ignores
whether the subject dies in remission or after relapse, whereas
the Aalen--Johansen uses the information from previous states 
through formula \eqref{eq:product}.

An exact weighting representation of the Aalen--Johansen is nonetheless available: \citet{datta2001} and \citet{yan2026} write it as an inverse-probability-weighted average whose weights depend on the state each subject occupies, in place of the single pooled censoring weight. Being an exact re-expression, it inherits the robustness of the Aalen--Johansen to non-Markov departures without explaining it.

A separate and more revealing question is whether the Aalen--Johansen instead belongs to the class of augmented IPCW estimators \citep{vdl2003}. Such an estimator supplements the IPCW estimator with a mean-zero augmentation built from a prediction of $I\{Y(t)=k\}$ from earlier history under a working model.
The augmentation can increase efficiency 
but incurs no bias whether the working model is correct or not.
Recently, \citet{maltzahn2026} show that the fully efficient augmentation uses the entire history just before $t$. Since the Aalen--Johansen carries the current state forward through a Markov model and is nonetheless robust to non-Markov departures,
it is natural to suspect that it is an IPCW estimator augmented under a working Markov model.

We show that this is indeed the case. The identity is exact in every sample and even holds for an arbitrary working censoring function. Set the working function to
the true censoring function, and three implications follow. First, the identity explains the robustness of the Aalen--Johansen under non-Markov departures. Second, it yields an intuitive influence function of the Aalen--Johansen that leads to a full efficiency analysis, especially in comparison with the semiparametric efficiency bound. Third, the efficiency analysis motivates using the Aalen--Johansen as the base for a practical, model-free augmentation.

\section{Aalen--Johansen as augmented IPCW}
\subsection{Observed data and the IPCW}
In addition to the notation defined in Section~\ref{sec:motivation},
let $D$ denote the time of entry into an absorbing state, such as death; $D=\infty$ if none exists. Observation through $D\wedge t$ determines $I\{Y(t)=k\}$, where $a\wedge b=\min(a,b)$. 
Let $C \in (0,\infty)$ be a continuously distributed censoring time, and assume that
$C\perp\!\!\perp\{Y(t):t\ge0\}$. 
Additionally,
let $X=C\wedge D$ denote the observed follow-up and $\delta=I(D\le C)$ the indicator of observed
absorption. Because $C>0$, $Y(0)$ is always observed.

For a random sample of $n$ subjects, we add a subscript $i$ for the quantities of subject $i$. With $N_{jki}(s)$ counting the observed $j\to k$ transitions by $s$, and $n_j(s)=\sum_{i=1}^n I\{X_i\ge s,\,Y_i(s-)=j\}$ the risk set in state $j$, the off-diagonal elements of $\mathrm d\widehat\Lambda(s)$ in \eqref{eq:product} are $\mathrm{d}\widehat\Lambda_{jk}(s)=\sum_{i=1}^n \mathrm{d}N_{jki}(s)/n_j(s)$ for $j\ne k$, zero for an empty risk set. So $I+\mathrm{d}\widehat\Lambda(s)$ holds the observed transition proportions within each origin state, including the proportion that stays in the diagonal elements. With $\widehat P(s,t)$ defined in \eqref{eq:product}, its $(j,k)$th entry $\widehat P_{jk}(s,t)$ is the fitted probability of occupying $k$ at $t$ when the working Markov model starts in $j$ at $s$.

To characterize the IPCW estimator, let $G^*$ be an arbitrary censoring survival function, with $G^*(0)=1$, $G^*(t)>0$, and no jumps at observed state changes or $t$. This can be the true function $G(t)=\prob(C>t)$,
an estimator, or even a faulty version of it.  The specific form of $G^*$ 
will not matter in the identity established in Theorem~\ref{thm:id}.
Under this $G^*$, the IPCW estimator is
\[
\widehat p_k^{\ipcw}(t;G^*)=n^{-1}\sum_{i=1}^n \frac{I\{C_i\ge D_i\wedge t,\,Y_i(t)=k\}}{G^*\{(D_i\wedge t)-\}}.
\]

\subsection{A finite-sample identity}

We now show that $\widehat p_k(t)$ is $\widehat p_k^{\ipcw}(t;G^*)$ plus an augmentation term.
Let $M_i^C(s;G^*)=N_i^C(s)-\int_0^s I(X_i\ge u)\dd\Lambda^*(u)$ denote
the censoring martingale under $G^*$,
where 
$N_i^C(s)=I(X_i\le s,\,\delta_i=0)$ 
and $\mathrm{d}\Lambda^*=-\mathrm{d}G^*/G^*_-$.

\begin{theorem}\label{thm:id}
For any such $G^*$ and every $t>0$,
\begin{equation}\label{eq:id}
 \widehat p_k(t)=\underbrace{n^{-1}\sum_{i=1}^n I(C_i\ge D_i\wedge t)\frac{I\{Y_i(t)=k\}}{G^*\{(D_i\wedge t)-\}}}_{\widehat p_k^{\ipcw}(t;G^*)}
 + 
 \underbrace{n^{-1}\sum_{i=1}^n\int_0^t
 \frac{\widehat P_{Y_i(s-),k}(s,t)}{G^*(s)}\dd M_i^C(s;G^*)}_{\text{Augmentation term}}.
\end{equation}
\end{theorem}

\begin{proof}[Proof of Theorem~\ref{thm:id}]
We sketch the idea; Appendix~A provides the
details. Both sides of \eqref{eq:id} equal the empirical proportion in
$k$ at $t=0$, and the full claim can be proved by induction over the
observed transition times. Between two such times, the left-hand side is
constant, while subject $i$'s contribution to the IPCW term changes by
$-I\{Y_i(t-)=k\}\,\mathrm dM_i^C(t;G^*)/G^*(t)$, cancelling the
augmentation term increment because $\widehat P(t,t)=I$. At a transition
time, the occupation vector updates by
$\widehat p(t)=\{I+\mathrm d\widehat\Lambda(t)\}^{\mathsf T}\widehat p(t-)$, by
\eqref{eq:product}, so the left-hand side, its $k$th entry, jumps
accordingly. The right-hand side jumps the same way: all subjects at
risk share the weight $1/G^*(t-)$, so the transitions move the IPCW term
as they move $\widehat p(t)$, and
$\widehat P(s,t)=\widehat P(s,t-)\{I+\mathrm d\widehat\Lambda(t)\}$ does
the same to the augmentation term.
\end{proof}

\subsection{Interpretation and implications}
Two immediate consequences follow from Theorem~\ref{thm:id}.
First, the left-hand side of \eqref{eq:id} does not involve $G^*$, so the identity is a family of decompositions of one quantity: as $G^*$ varies, the two terms on the right change but their sum does not. The Aalen--Johansen thus needs no censoring estimator; $G^*$ is a device of the analysis. At $G^*=G$, the weighted term is consistent and the augmentation term has asymptotic mean zero, which recovers the known 
robustness property of the Aalen--Johansen. Efficiency analysis requires more effort and is deferred to Section~\ref{sec:asymp}.

Second, we can use \eqref{eq:id} to recover the numerical equivalence
between $\widehat p_k(t)$ and the IPCW estimator in the univariate
and competing risks settings noted in Section~\ref{sec:motivation}.
Indeed, take $G^*=\widehat G$, the Kaplan--Meier estimator of $G$. We can then re-express the augmentation term in \eqref{eq:id}
as a weighted integral of the difference between $\widehat P_{Y_i(s-),k}(s,t)$
and its average over the origin states $0,1,\ldots,K$ of the subjects at risk.
This difference vanishes if all subjects at risk occupy the same state,
as is the case in Figure~\ref{fig:models}(a) and (b), where there is a single transient state.
\begin{proposition}\label{prop:km}
Let $\widehat G(s)=\prod_{u\le s}\{1-\sum_{i=1}^n\mathrm{d}N_i^C(u)/r(u)\}$ denote the Kaplan--Meier estimator of $G$, where $r(u)=\sum_{j=0}^K n_j(u)$ is the number of subjects at risk at $u$, and let $\overline P_k(s,t)=\sum_{j=0}^K\{n_j(s)/r(s)\}\widehat P_{jk}(s,t)$ denote the average of $\widehat P_{Y_i(s-),k}(s,t)$ over the subjects at risk at $s$. If $\widehat G(t)>0$, then
\begin{equation}\label{eq:imbalance}
 \widehat p_k(t)-\widehat p_k^{\ipcw}(t;\widehat G)
 =n^{-1}\sum_{i=1}^n\int_0^t\frac{\widehat P_{Y_i(s-),k}(s,t)-\overline P_k(s,t)}{\widehat G(s)}\dd N_i^C(s).
\end{equation}
Consequently,
\begin{itemize}
\item[(i)] for given $k$ and $t$, $\widehat p_k(t)=\widehat p_k^{\ipcw}(t;\widehat G)$ if, at every censoring time before $t$, all subjects at risk occupy the same state;
\item[(ii)] $\widehat p_k(t)=\widehat p_k^{\ipcw}(t;\widehat G)$ for all $k$ and all $t$ with $\widehat G(t)>0$ if and only if, at every censoring time, all subjects at risk occupy the same state.
\end{itemize}
\end{proposition}
The proof is relegated to Appendix~B.
In particular, (ii) shows that the univariate and competing risks settings,
with a single transient state, are the only ones in which
a global equivalence between the Aalen--Johansen and the IPCW holds in every sample.

\section{Asymptotic analysis}\label{sec:asymp}
The representation of $\widehat p_k(t)$ in \eqref{eq:id} makes it
amenable to asymptotic efficiency analysis.
Indeed, given the uniform convergence of $\widehat P(s,t)$ to $P(s,t)$,
we can 
replace $\widehat P$ by $P$ in \eqref{eq:id} with only a $o_p(n^{-1/2})$ change to $\widehat p_k(t)$.
Setting $G^*=G$, we obtain the following linearized form of $\widehat p_k(t)$.
\begin{proposition}\label{prop:lin}
Suppose $\Lambda(t)$ is finite, with $\mathrm{d}\Lambda(t)$ defined in the second equality of \eqref{eq:markov}. Then
\begin{equation}\label{eq:ajif}
\begin{split}
 n^{1/2}\{\widehat p_k(t)-p_k(t)\}
 =n^{-1/2}\sum_{i=1}^n\bigg[&\frac{I\{C_i\ge D_i\wedge t,\,Y_i(t)=k\}}{G\{(D_i\wedge t)-\}}-p_k(t)\\
 &+\int_0^t \frac{P_{Y_i(s-),k}(s,t)}{G(s)}\dd M_i^C(s;G)\bigg]+o_p(1).
\end{split}
\end{equation}
\end{proposition}
The full proof is given in Appendix~C. 

The geometry of the influence function in \eqref{eq:ajif} plays a key
role in the efficiency analysis of $\widehat p_k(t)$.
To investigate this geometry, we need to first understand what
$P_{jk}(s,t)$ truly is in general.
Analytically, $P(s, t)$ is the product integral of $\mathrm{d}\Lambda(t)$ in \eqref{eq:prodint}. The entries of $\mathrm{d}\Lambda(t)$ are defined in the second equality of \eqref{eq:markov} and can be viewed as average intensities across earlier history,
that is, $\mathrm{d}\Lambda_{jk}(t)=E\left[\prob\{Y(t+\mathrm d t) = k\mid Y(t-)=j, \mathcal H(t-)\}\mid Y(t-)=j\right]$ in non-Markov scenarios.
We might expect $P_{jk}(s,t)$ to be an average transition probability
in a similar sense, but that is not the case \citep{putter2018}.

To avoid confusion, we define the quantities of interest.
\begin{definition}\label{def:trans}
For states $j$ and $k$ and times $s\le t$,
\begin{itemize}
\item[(i)] The transition probability given the history is 
$$Q_{jk}\{s,t;\mathcal H(s-)\}=\prob\{Y(t)=k\mid Y(s-)=j,\,\mathcal H(s-)\};$$
\item[(ii)] The average transition probability is 
$$q_{jk}(s,t)=\prob\{Y(t)=k\mid Y(s-)=j\}=E[Q_{jk}\{s,t;\mathcal H(s-)\}\mid Y(s-)=j];$$
\item[(iii)] The Markov surrogate of the transition probability is $P_{jk}(s,t)$,
which is the asymptotic limit of $\widehat P_{jk}(s,t)$ and the $(j,k)$th entry of the product integral as given in \eqref{eq:prodint}.
\end{itemize}
\end{definition}

The three quantities coincide under \eqref{eq:markov} when $Y(s)=Y(s-)$, as holds almost surely at a censoring time $s$.
In general, $Q_{jk}\{s,t;\mathcal H(s-)\}\neq q_{jk}(s,t)$ because
$Y(\cdot)$ need not be Markov at $s$. The discrepancy between
$q_{jk}(s,t)$ and $P_{jk}(s,t)$ can be seen from the definition
of the latter in \eqref{eq:prodint}.
As $u$ increases from $s$ to $t$ in the product integral,
$\mathrm d\Lambda(u)$ does not retain the conditioning event
$Y(s-)=j$ but conditions on the current state $Y(u-)$ alone, as
the Markov assumption would allow. This incurs bias if the process
is non-Markov over $(s,t]$. In summary,
\begin{equation}\label{eq:ineq}
  Q_{jk}\{s,t;\mathcal H(s-)\}
  \;\underset{\text{non-Markovness at }s}{\neq}\;
  q_{jk}(s,t)
  \;\underset{\text{non-Markovness over }(s,t]}{\neq}\;
  P_{jk}(s,t).
\end{equation}
These relationships are important because $Q_{jk}$ and $q_{jk}$
play a key role in benchmarking the efficiency of $\widehat p_k(t)$,
which uses $P_{jk}$ in the augmentation.

To start, \citet{maltzahn2026} show that the efficient influence function
of $p_k(t)$ is the influence function in \eqref{eq:ajif} with
$P_{jk}(s,t)$ replaced by $Q_{jk}\{s,t;\mathcal H(s-)\}$;
that is, for a generic subject,
\begin{equation}\label{eq:eff}
\phi_Q^{\mathrm{eff}}(t)=\frac{I\{C\ge D\wedge t,\,Y(t)=k\}}{G\{(D\wedge t)-\}}-p_k(t)+\int_0^t \frac{Q_{Y(s-),k}\{s,t;\mathcal H(s-)\}}{G(s)}\dd M^C(s;G).
\end{equation}
This is obtained by
$\phi_Q^{\mathrm{eff}}(t)=\phi^{\ipcw}(t)-\Pi\{\phi^{\ipcw}(t)\mid\mathcal S\}$,
where $\phi^{\ipcw}(t)=I\{C\ge D\wedge t,\,Y(t)=k\}/G\{(D\wedge t)-\}-p_k(t)$ is
the centred IPCW contribution, $\Pi(\cdot\mid\mathcal S)$ denotes projection onto $\mathcal S$, and
 $\mathcal S=\big\{\int_0^t h\{s;\mathcal H(s-)\}\dd M^C(s;G): h\text{ is square-integrable}\big\}$
is the nuisance tangent space for censoring.
To achieve \eqref{eq:eff}, we would need an augmentation term that approximates
$Q_{jk}\{s,t;\mathcal H(s-)\}$, which depends on the full history $\mathcal H(s-)$.

For a simpler augmentation, at a possible price of full efficiency,
we can instead project $\phi^{\ipcw}(t)$ onto a proper subspace of $\mathcal S$,
such as martingale integrals of functions of the current state:
\[
 \mathcal S_0=\left\{\int_0^t a\{s; Y(s-)\}\dd M^C(s;G): a\text{ is square-integrable}\right\}\subset\mathcal S.
\]
Let $\phi_q(t)=\phi^{\ipcw}(t)-\Pi\{\phi^{\ipcw}(t)\mid\mathcal S_0\}$ denote the resulting influence function.
\begin{proposition}\label{prop:proj}
 We have that
 \begin{equation*}
   \phi_q(t)=\frac{I\{C\ge D\wedge t,\,Y(t)=k\}}{G\{(D\wedge t)-\}}-p_k(t)+\int_0^t \frac{q_{Y(s-),k}(s,t)}{G(s)}\dd M^C(s;G).
 \end{equation*}
 Thus, the current-state augmentation uses $q_{jk}(s,t)$ rather than $P_{jk}(s,t)$ in general.
\end{proposition}
The proof is given in Appendix~C.

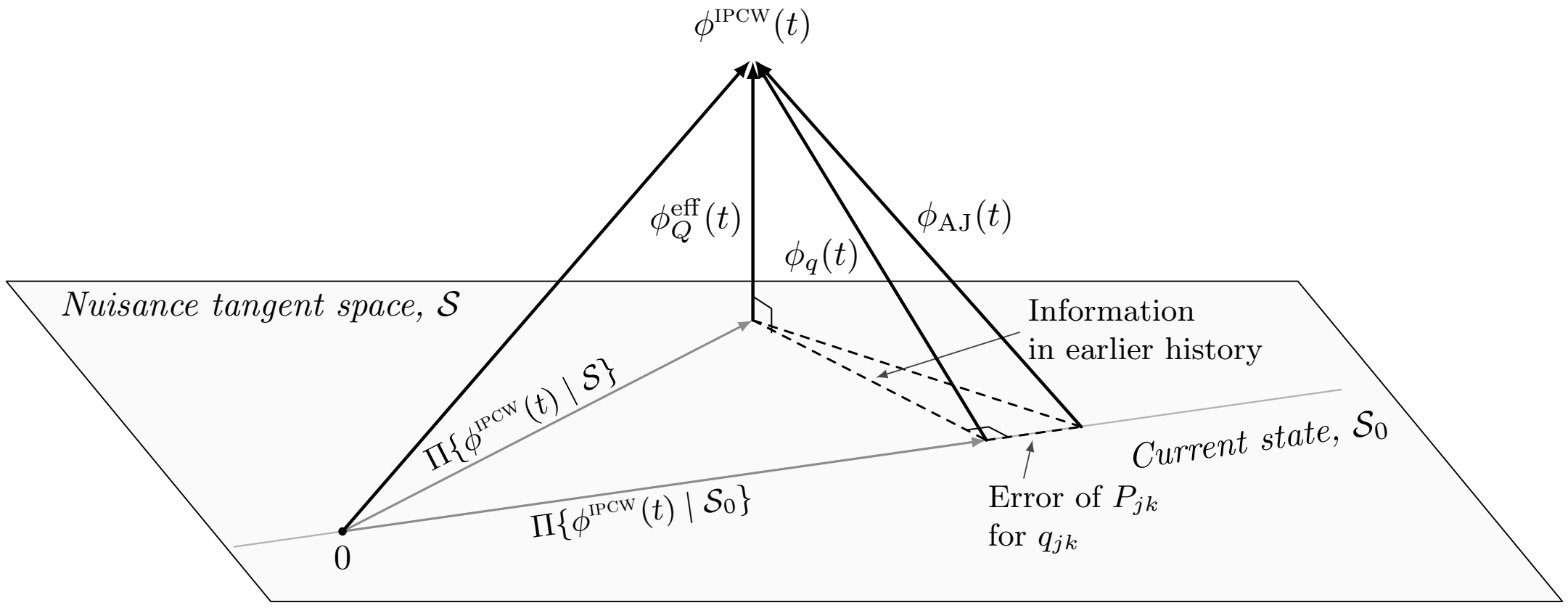
\begin{figure}[tb]
\centering
\setlength{\figwidth}{.86\linewidth}
\resizebox{.86\linewidth}{!}{
\begin{tikzpicture}[x=1cm,y=1cm,>=latex,line cap=round,line join=round,
    every node/.style={text=black,inner sep=1pt,font=\small,scale=1.25},
    note/.style={font=\footnotesize},
    influence/.style={->,line width=1.15pt,shorten >=1.5pt},
    projection/.style={->,black!45,line width=.8pt}]
  \coordinate (origin) at (2.05,.35);
  \coordinate (full)   at (7.2,3.0);       
  \coordinate (ipcw)   at (7.2,6.3);
  \coordinate (state)  at (10.13,1.495);     
  \coordinate (aj)     at (11.32,1.663);    

  \draw[line width=.49pt,fill=black!2]
    (1.15,-.53) -- (17.36,-.53) -- (14.04,3.49) -- (-2.17,3.49) -- cycle;
  \draw[black!30,line width=.6pt] (.69,.157) -- (14.58,2.13);
  \node[anchor=west,font=\small\itshape] at (-1.55,3.17) {Nuisance tangent space, $\mathcal S$};
  \node[anchor=north east,rotate=8,font=\small\itshape] at (15.2,1.95) {Current state, $\mathcal S_0$};

  \node[anchor=east] at (7.13,4.25) {$\phi_Q^{\mathrm{eff}}(t)$};
  \node[anchor=east] at (8.6,3.8) {$\phi_q(t)$};
  \node[anchor=west] at (9.2,4.3) {$\phi_{\mathrm{AJ}}(t)$};

  \draw[line width=.52pt] (10.407,1.534) -- (10.158,1.662) -- (9.881,1.623);
  \draw[line width=.52pt] (7.2,3.3) -- (7.434,3.146) -- (7.434,2.846);

  \draw[projection] (origin) -- (state)
    node[note,pos=.46,sloped,below,text=black,yshift=-1pt] {$\Pi\{\phi^{\ipcw}(t)\mid\mathcal S_0\}$};
  \draw[projection] (origin) -- (full)
    node[note,pos=.46,sloped,above,text=black] {$\Pi\{\phi^{\ipcw}(t)\mid\mathcal S\}$};
  \draw[dashed,line width=.76pt] (full) -- (state);
  \draw[dashed,line width=.76pt] (state) -- (aj);
  \draw[dashed,line width=.76pt] (full) -- (aj);
  \node[note,anchor=west,align=left] (historylabel) at (10.6,2.85) {Information\\in earlier history};
  \draw[->,black!72,line width=.49pt] ([xshift=-1.5pt]historylabel.west) -- (8.74,2.28);
  \node[note,anchor=north west,align=left] (errorlabel) at (10.1,.98) {Error of $P_{jk}$\\for $q_{jk}$};
  \draw[->,black!72,line width=.49pt] (10.6,1.02) -- (10.71,1.5);
  \draw[influence] (origin) -- (ipcw);
  \draw[influence] (full) -- (ipcw);
  \draw[influence] (state) -- (ipcw);
  \draw[influence] (aj) -- (ipcw);
  \fill (origin) circle[radius=1.5pt];
  \node[anchor=north] at (2.05,.22) {$0$};
  \node[anchor=south] at (7.2,6.42) {$\phi^{\ipcw}(t)$};
\end{tikzpicture}}
\caption{Geometry of the influence functions. The dashed-line, right-angled triangle explains the Pythagorean-like decomposition of the efficiency shortfall of $\phi_{\mathrm{AJ}}(t)$ relative to $\phi_Q^{\mathrm{eff}}(t)$, as formulated in \eqref{eq:gap} of Proposition~\ref{prop:var}. 
In a Markov model, the triangle collapses into a point, so that $\phi_{\mathrm{AJ}}(t)=\phi_{q}(t)=\phi_Q^{\mathrm{eff}}(t)$.}
\label{fig:projection}
\end{figure}

Now, let $\phi_{\mathrm{AJ}}(t)$ denote the influence function
of $\widehat p_k(t)$ in \eqref{eq:ajif}. Under a Markov model,
$\phi_{\mathrm{AJ}}(t)=\phi_Q^{\mathrm{eff}}(t)$ because $P_{jk}(s,t)=Q_{jk}\{s,t;\mathcal H(s-)\}$, and $\widehat p_k(t)$ is hence fully efficient.
In general, the efficiency loss can be attributed to a two-step
approximation of $Q_{jk}$ by $P_{jk}$. First, using $q_{jk}(s,t)$
in place of $Q_{jk}\{s,t;\mathcal H(s-)\}$ loses possible information
in $\mathcal H(s-)$ if $Y(\cdot)$ is non-Markov at $s$;
second, using $P_{jk}(s,t)$ in place of $q_{jk}(s,t)$ departs from
the best current-state augmentation if $Y(\cdot)$ is non-Markov over $(s,t]$,
as summarised in \eqref{eq:ineq}. The two steps are orthogonal, since
$\phi_q(t)-\phi_Q^{\mathrm{eff}}(t)$ is orthogonal to $\mathcal S_0$ while
$\phi_{\mathrm{AJ}}(t)-\phi_q(t)$ lies in it, so that their contributions to the variance add.
This is illustrated in Figure~\ref{fig:projection} and quantified in the next proposition. 

\begin{proposition}\label{prop:var}
Let $\Lambda_C=-\log G$. Under the conditions of Proposition~\ref{prop:lin}, 
\begin{equation}\label{eq:gap}
\begin{split}
 \mathrm{var}\{\phi_{\mathrm{AJ}}(t)\}-\mathrm{var}\{\phi_{Q}^{\mathrm{eff}}(t)\}
 =\int_0^t&E
 \Big(\underbrace{\var[Q_{Y(s-),k}\{s,t;\mathcal H(s-)\}\mid Y(s-)]}_{\text{Individual variation of $Q_{jk}$ from mean $q_{jk}$}}\\
 &+\underbrace{\{P_{Y(s-),k}(s,t)-q_{Y(s-),k}(s,t)\}^2}_{\text{Approximation error of $P_{jk}$ for mean $q_{jk}$}}\Big)
 \frac{\mathrm{d}\Lambda_C(s)}{G(s)}.
\end{split}
\end{equation}
\end{proposition}
Appendix~C provides a general efficiency
comparison for arbitrary augmentation terms, with those in \eqref{eq:gap}
and the IPCW as special cases. In the simulations of Appendix~E under
Markov and non-Markov models, the variance of the IPCW with Kaplan--Meier
weights exceeds that of the Aalen--Johansen by up to $34\%$, suggesting
that the latter is a better choice of a nonparametric base estimator.

\section{Model-free least-squares augmentation of the Aalen--Johansen}

To attain \eqref{eq:eff}, one would replace $Q_{jk}\{s,t;\mathcal H(s-)\}$ by a fitted full-history prognosis and augment the IPCW accordingly \citep{maltzahn2026}. This requires a model for the entire process, from which the prognosis must typically be obtained by simulation outside the Markov case. If the model is wrong, the estimator retains only the consistency of the IPCW, with no assurance of improving on $\widehat p_k(t)$. 

We instead propose a model-free least-squares augmentation of the Aalen--Johansen. By \eqref{eq:ajif} and \eqref{eq:eff}, the influence function of $\widehat p_k(t)$ differs with the efficient one by the augmentation with integrand $Q_{Y(s-),k}\{s,t;\mathcal H(s-)\}-P_{Y(s-),k}(s,t)$. Let
\begin{equation}\label{eq:delta}
 \Delta_{jk}\{s,t;\mathcal H(s-)\}=Q_{jk}\{s,t;\mathcal H(s-)\}-q_{jk}(s,t)
\end{equation}
denote what earlier history adds to the current state, so that $Q_{jk}\{s,t;\mathcal H(s-)\}-P_{jk}(s,t)=\Delta_{jk}\{s,t;\mathcal H(s-)\}+\{q_{jk}(s,t)-P_{jk}(s,t)\}$. The two terms are orthogonal and correspond to those in \eqref{eq:gap}. By Proposition~\ref{prop:var}, approximating $\Delta_{jk}$ thus reduces the first term of the variance gap without affecting the second. By Definition~\ref{def:trans}, $E[\Delta_{jk}\{s,t;\mathcal H(s-)\}\mid Y(s-)=j]=0$.

A working version of $\Delta_{jk}$ should share this property. Let $h\{s;\mathcal H(s-)\}$ be a prespecified score summarizing the history $\mathcal H(s-)$, such as the time spent in the current state, $h\{s;\mathcal H(s-)\}=s-\sup\{u<s:Y(u)\ne Y(s-)\}$ with $\sup\emptyset=0$, which accounts for semi-Markov dependence. 
To satisfy the conditional mean-zero property, 
we centre it within the current state by
\begin{align*}
     e_i(s)&=h\{s;\mathcal H_i(s-)\}-\bar h_{Y_i(s-)}(s),\\
     &\qquad\mbox{where }
 \bar h_j(s)=n_j(s)^{-1}\sum_{i=1}^n I\{X_i\ge s,\,Y_i(s-)=j\}h\{s;\mathcal H_i(s-)\}.
\end{align*}

Similarly to least-squares augmentation in covariate adjustment for censored outcomes \citep{lu2008,zhang2008}, we approximate $\Delta_{Y_i(s-),k}\{s,t;\mathcal H_i(s-)\}$ by the linear working model $\beta e_i(s)$ and choose $\beta$ by regression. Because $\Delta_{jk}$ is the mean of $I\{Y(t)=k\}$ given the history, subtracted by its mean within the state, and $e_i(s)$ is already centred within the state, this amounts to regressing $I\{Y_i(t)=k\}$ on $e_i(s)$ among the subjects at risk at $s$. Under censoring, the response is replaced by $o_i(s)=I\{C_i\ge D_i\wedge t,\,Y_i(t)=k\}\widehat G(s-)/\widehat G\{(D_i\wedge t)-\}$, which is weighted for censoring after $s$ and has the same mean given the history. Let $\widehat\beta$ be the least-squares slope of $o_i(s)$ on $e_i(s)$, pooled over the censoring times $s$, where the augmentation acts. Adding the fitted augmentation $n^{-1}\sum_{i=1}^n\int_0^t\widehat\beta e_i(s)\widehat G(s)^{-1}\dd M_i^C(s;\widehat G)$ to the Aalen--Johansen, we obtain
\begin{equation}\label{eq:adj}
 \widehat p_k^{\rm aug}(t)=\widehat p_k(t)+\widehat\beta\,\widehat c_h(t),\qquad
 \widehat c_h(t)=n^{-1}\sum_{i=1}^n I(C_i<D_i\wedge t)\,\widehat G(C_i)^{-1}e_i(C_i).
\end{equation}
The details of this procedure are given in Appendix~D.

\begin{proposition}\label{prop:adj}
The estimator in \eqref{eq:adj} has the following properties.
\begin{itemize}
\item[(i)] If $h\{s;\mathcal H(s-)\}=a\{s; Y(s-)\}$ for some function $a$, then $\widehat p_k^{\rm aug}(t)=\widehat p_k(t)$ in every sample.
\item[(ii)] For a general score $h$, under the regularity conditions of Appendix~D,
\[
 \limsup_{n\to\infty}\,n\big[\var\{\widehat p_k^{\rm aug}(t)\}-\var\{\widehat p_k(t)\}\big]\le0.
\]
\end{itemize}
\end{proposition}

Thus, a score carrying no information beyond the current state leaves the Aalen--Johansen unchanged, and any other cannot increase its asymptotic variance. Appendix~D gives the proof.
In the simulations of Appendix~E, with $h$ the time spent in the current state, the variance falls by $3$--$16\%$ when the transition rate increases with that time, and changes by at most $1\%$ otherwise.
No part of the process need be modelled or simulated.

\section*{Acknowledgement}
This research was supported by U.S. National Institutes of Health grant R01HL149875.

\bibliographystyle{abbrvnat}
\bibliography{references}

\clearpage
\appendix
\setcounter{equation}{0}
\setcounter{figure}{0}
\setcounter{table}{0}
\renewcommand{\theequation}{A\arabic{equation}}
\renewcommand{\thefigure}{A\arabic{figure}}
\renewcommand{\thetable}{A\arabic{table}}
\renewcommand{\theHequation}{A\arabic{equation}}
\renewcommand{\theHfigure}{A\arabic{figure}}
\renewcommand{\theHtable}{A\arabic{table}}

\begin{center}
{\Large\bfseries Appendix}
\end{center}

\section{Proof of Theorem~1}

Let $W_i(t)=I(C_i\ge D_i\wedge t)/G^*\{(D_i\wedge t)-\}$ denote the weight of subject $i$, so that $\widehat p_k^{\ipcw}(t;G^*)=n^{-1}\sum_{i=1}^n W_i(t)I\{Y_i(t)=k\}$, and let
\[
 \widehat U_k(t)=n^{-1}\sum_{i=1}^n\int_0^t G^*(s)^{-1}\widehat P_{Y_i(s-),k}(s,t)\dd M_i^C(s;G^*)
\]
denote the augmentation term, where $M_i^C(s;G^*)=N_i^C(s)-\int_0^s I(X_i\ge u)\dd\Lambda^*(u)$, $N_i^C(s)=I(X_i\le s,\,\delta_i=0)$ and $\mathrm{d}\Lambda^*=-\mathrm{d}G^*/G^*_-$, as in the main text. Theorem~1 states that
\begin{equation}\label{eq:claim}
 \widehat p_k(t)=\widehat p_k^{\ipcw}(t;G^*)+\widehat U_k(t)\quad(k=0,\ldots,K).
\end{equation}

We first show that
\begin{equation}\label{eq:weight}
 W_i(t)=1-\int_0^t G^*(s)^{-1}\dd M_i^C(s;G^*).
\end{equation}
Since $\mathrm{d}(1/G^*)=\mathrm{d}\Lambda^*/G^*$ and $G^*(0)=1$,
\[
 \int_0^t I(X_i\ge s)G^*(s)^{-1}\dd\Lambda^*(s)
 =\int_0^{X_i\wedge t}\dd\{1/G^*(s)\}
 =G^*(X_i\wedge t)^{-1}-1,
\]
so that the right-hand side of \eqref{eq:weight} equals $G^*(X_i\wedge t)^{-1}-N_i^C(t)G^*(X_i)^{-1}$. If $C_i\ge D_i\wedge t$, then $N_i^C(t)=0$ and $X_i\wedge t=D_i\wedge t$, and this expression is $G^*(D_i\wedge t)^{-1}=W_i(t)$, because $G^*$ has no jumps at observed state changes or $t$. If $C_i<D_i\wedge t$, then $N_i^C(t)=1$ and $X_i\wedge t=C_i$, and this expression is zero, as is $W_i(t)$. The event $C_i=t<D_i$, on which $N_i^C(t)=1$ while $W_i(t)>0$, has probability zero.

Let $0=t_0<t_1<\cdots<t_m$ denote the observed transition times. We show by induction on $l$ that \eqref{eq:claim} holds on $[0,t_l]$. At $t_0=0$, $W_i(0)=1$ and $\widehat U_k(0)=0$, so that both sides of \eqref{eq:claim} are the empirical proportion in $k$.

Suppose that \eqref{eq:claim} holds at $t_l$, and let $t_l\le t<t_{l+1}$. No transition is observed in $(t_l,t]$, so that $\widehat p_k(t)=\widehat p_k(t_l)$ and each subject under observation keeps its state. By \eqref{eq:weight},
\[
 \widehat p_k^{\ipcw}(t;G^*)-\widehat p_k^{\ipcw}(t_l;G^*)
 =-n^{-1}\sum_{i=1}^n\int_{(t_l,t]}G^*(s)^{-1}I\{Y_i(s-)=k\}\dd M_i^C(s;G^*).
\]
For the same reason, $\widehat P(s,t)=\widehat P(s,t_l)$ for $s\le t_l$ and $\widehat P(s,t)$ is the identity matrix for $t_l<s\le t$, so that $\widehat U_k(t)-\widehat U_k(t_l)$ is the negative of the last display. Hence \eqref{eq:claim} holds on $[t_l,t_{l+1})$.

Now let $t=t_{l+1}$. By the product formula of the main text, $\widehat p_k(t)=\widehat p_k(t-)+\sum_{j=0}^K\widehat p_j(t-)\,\mathrm{d}\widehat\Lambda_{jk}(t)$. It suffices to show that the two terms on the right of \eqref{eq:claim} change in the same way, since \eqref{eq:claim} holds at $t-$ for every state. Because $\widehat P(s,t)=\widehat P(s,t-)\{I+\mathrm{d}\widehat\Lambda(t)\}$ for $s<t$, and $M_i^C(\cdot;G^*)$ has no jump at $t$, we have $\widehat U_k(t)=\widehat U_k(t-)+\sum_{j=0}^K\widehat U_j(t-)\,\mathrm{d}\widehat\Lambda_{jk}(t)$. For the weighted term, censoring is untied with observed state changes and $G^*$ has no jumps there, so that $W_i$ does not jump at $t$, and a subject whose state changes at $t$ is at risk with $W_i(t-)=G^*(t-)^{-1}$. Since $\mathrm{d}I\{Y_i(t)=k\}=\sum_{j\ne k}\{\mathrm{d}N_{jki}(t)-\mathrm{d}N_{kji}(t)\}$,
\begin{equation}\label{eq:netflow}
 \widehat p_k^{\ipcw}(t;G^*)-\widehat p_k^{\ipcw}(t-;G^*)
 =n^{-1}G^*(t-)^{-1}\sum_{i=1}^n\sum_{j\ne k}\{\mathrm{d}N_{jki}(t)-\mathrm{d}N_{kji}(t)\}.
\end{equation}
For a transient state $j$, every subject with $X_i\ge t$ and $Y_i(t-)=j$ has $D_i\ge t$ and hence weight $G^*(t-)^{-1}$ just before $t$, so that $\widehat p_j^{\ipcw}(t-;G^*)=n_j(t)/\{nG^*(t-)\}$. For $j\ne k$, the risk set therefore cancels:
\[
 \widehat p_j^{\ipcw}(t-;G^*)\,\mathrm{d}\widehat\Lambda_{jk}(t)
 =\frac{n_j(t)}{nG^*(t-)}\,\frac{\sum_{i=1}^n\mathrm{d}N_{jki}(t)}{n_j(t)}
 =n^{-1}G^*(t-)^{-1}\sum_{i=1}^n\mathrm{d}N_{jki}(t).
\]
For an absorbing state $j$, no exit is observed, so that $\mathrm{d}\widehat\Lambda_{jk}(t)=0$ for all $k$ and the weights frozen at absorption do not enter. Since $\mathrm{d}\widehat\Lambda_{kk}(t)=-\sum_{j\ne k}\mathrm{d}\widehat\Lambda_{kj}(t)$, the right-hand side of \eqref{eq:netflow} is $\sum_{j=0}^K\widehat p_j^{\ipcw}(t-;G^*)\,\mathrm{d}\widehat\Lambda_{jk}(t)$. Hence \eqref{eq:claim} holds at $t_{l+1}$, which completes the induction.

\section{Proof of Proposition~1}

Set $G^*=\widehat G$ in Theorem~1. The conditions on $G^*$ hold because $\widehat G(0)=1$, $\widehat G(t)>0$ and $\widehat G$ jumps only at censoring times, which are untied with the observed state changes and with $t$. Hence
\begin{equation}\label{eq:kmid}
 \widehat p_k(t)-\widehat p_k^{\ipcw}(t;\widehat G)
 =n^{-1}\sum_{i=1}^n\int_0^t\widehat G(s)^{-1}\widehat P_{Y_i(s-),k}(s,t)\dd M_i^C(s;\widehat G),
\end{equation}
where $M_i^C(s;\widehat G)=N_i^C(s)-\int_0^s I(X_i\ge u)\dd\Lambda^*(u)$ and $\mathrm{d}\Lambda^*=-\mathrm{d}\widehat G/\widehat G_-$. By the definition of the Kaplan--Meier estimator, $\mathrm{d}\Lambda^*(s)=r(s)^{-1}\sum_{l=1}^n\mathrm{d}N_l^C(s)$. The compensator part of \eqref{eq:kmid} is therefore
\begin{align*}
 &n^{-1}\sum_{l=1}^n\int_0^t\widehat G(s)^{-1}\left\{r(s)^{-1}\sum_{i=1}^n I(X_i\ge s)\widehat P_{Y_i(s-),k}(s,t)\right\}\dd N_l^C(s)\\
 &\qquad=n^{-1}\sum_{l=1}^n\int_0^t\widehat G(s)^{-1}\overline P_k(s,t)\dd N_l^C(s),
\end{align*}
because the average in braces equals $\sum_{j=0}^K\{n_j(s)/r(s)\}\widehat P_{jk}(s,t)=\overline P_k(s,t)$. Subtracting this from the part of \eqref{eq:kmid} integrated against $N_i^C$ gives the display in Proposition~1. If all subjects at risk at a censoring time $s$ occupy the same state $j$, then $\overline P_k(s,t)=\widehat P_{jk}(s,t)$ and the integrand vanishes at $s$ for every $k$ and $t$, which gives (i) and the sufficiency in (ii). For the necessity in (ii), suppose that the condition fails, and let $s$ denote the first censoring time at which the subjects at risk occupy more than one state. Take $t>s$ before the next observed transition or censoring time, and let $k$ denote the state of the subject censored at $s$. The censoring times before $s$ contribute nothing to the display in Proposition~1. No transition is observed in $(s,t]$, so that $\widehat P(s,t)$ is the identity matrix and the integrand at $s$ is $\{1-n_k(s)/r(s)\}/\widehat G(s)$, which is positive because some subject at risk occupies another state. Hence $\widehat p_k(t)\ne\widehat p_k^{\ipcw}(t;\widehat G)$.

\section{Projection using the current state or the full history}

Let $Y(t)$ be the state at time $t$ and $p_k(t)=\pr\{Y(t)=k\}$. Suppose that censoring is independent of the life history $\{Y(t):t\ge0\}$, with survival function $G$ positive through $t$. Under the observation scheme of the main text, follow-up through $D\wedge t$, the earlier of absorption and $t$, determines $I\{Y(t)=k\}$. Let $Q_{jk}\{s,t;\mathcal H(s-)\}$, $q_{jk}(s,t)$ and $P_{jk}(s,t)$ be as in Definition~1 of the main text. Let $C$ be the censoring time. Let $X=C\wedge D$ denote the observed follow-up time and $\Lambda_C$ the common censoring hazard, and let $M^C(s;G)$ be the censoring martingale of the main text for a generic subject, with compensator $\int_0^{s}I(X\ge u)\dd\Lambda_C(u)$.

\subsection{Asymptotic linearity of the Aalen--Johansen}
We first show that replacing $\widehat P$ by $P$ in the identity of Theorem~1 has negligible first-order effect. Let $y_j(u)=\pr\{X\ge u,\,Y(u-)=j\}$. Because $C$ is independent of the life history, the transition rate $\mathrm{d}\Lambda_{jl}(u)$ defined by the second expression in display~(1) of the main text equals $\mathrm{d}\E\{N_{jl}(u)\}/y_j(u)$, where $N_{jl}(u)$ counts the observed $j\to l$ transitions of a subject by $u$.

Suppose that $\Lambda_{jl}(t)<\infty$ for each transient state $j$ and each $l\ne j$. We first show that $\sup_{s\le t}|\widehat\Lambda_{jl}(s)-\Lambda_{jl}(s)|\to0$ in probability. For $\eta>0$, let $B_\eta=\{u\le t:y_j(u)<\eta\}$. Given that subject $i$ is at risk in $j$ at $u$, the number $n_j(u)-1$ of other such subjects is binomial and independent of subject $i$, so that $\E\{1/n_j(u)\}\le1/\{ny_j(u)\}$ given that event. Hence $\E\{\int_{B_\eta}\mathrm{d}\widehat\Lambda_{jl}(u)\}\le\int_{B_\eta}\mathrm{d}\Lambda_{jl}(u)$, which tends to zero as $\eta\to0$ because $\Lambda_{jl}(t)<\infty$. Outside $B_\eta$, $y_j$ is at least $\eta$, and $n^{-1}n_j(u)$ and $n^{-1}\sum_{i=1}^n N_{jli}(u)$ converge uniformly in probability to $y_j(u)$ and $\E\{N_{jl}(u)\}$ by the Glivenko--Cantelli theorem, the expected number of transitions being finite because $\E\{N_{jl}(t)\}\le\Lambda_{jl}(t)$. Hence the part of $\widehat\Lambda_{jl}(s)$ accumulated outside $B_\eta$ converges uniformly to that of $\Lambda_{jl}(s)$, and letting $\eta\to0$ gives the claim.

For a vector $h=\{h_j\}$ of functions on $[0,t]$ indexed by the transient states, let
\[
 f_h=\int_0^t G(s)^{-1}h_{Y(s-)}(s)\dd M^C(s;G),
\]
and let $f_{h,i}$ denote its value for subject $i$. With $G^*=G$, the change in the augmentation term is $R_n=n^{-1}\sum_{i=1}^n(f_{\widehat h,i}-f_{h_0,i})$, where $\widehat h_j(s)=\widehat P_{jk}(s,t)$ and $h_{0j}(s)=P_{jk}(s,t)$. Because $C$ is independent of the life history, $M^C(\cdot;G)$ is a martingale given the path of $Y$, so that $\E(f_h)=0$ for every fixed $h$. Let $\mathcal F$ denote the class of $h$ with each $h_j$ bounded by one in absolute value and of total variation at most $V$. Each $f_h$ is the sum of $-\int_0^{X\wedge t}G(s)^{-1}h_{Y(s-)}(s)\dd\Lambda_C(s)$ and, for a subject censored before $D\wedge t$, $G(X)^{-1}h_{Y(X-)}(X)$. As uniformly bounded functions of bounded variation form a Donsker class, so does $\{f_h:h\in\mathcal F\}$. Since $\widehat P(s-,t)-\widehat P(s,t)=\mathrm{d}\widehat\Lambda(s)\widehat P(s,t)$ and the entries of $\widehat P$ lie in $[0,1]$, the total variation of $\widehat h_j$ on $[0,t]$ is at most $2\sum_{l\ne j}\widehat\Lambda_{jl}(t)$, so that $\widehat h\in\mathcal F$ with probability tending to one when $V>2\max_j\sum_{l\ne j}\Lambda_{jl}(t)$. By the continuity of product integration, $\sup_{s\le t}|\widehat h_j(s)-h_{0j}(s)|\to0$ in probability, so that $\E\{(f_h-f_{h_0})^2\}$ evaluated at $h=\widehat h$ tends to zero in probability. The asymptotic equicontinuity of the empirical process indexed by a Donsker class then gives
\[
 R_n=n^{-1}\sum_{i=1}^n\{f_{\widehat h,i}-f_{h_0,i}\}-\E(f_h-f_{h_0})\big|_{h=\widehat h}=o_p(n^{-1/2}).
\]

By Theorem~1 with $G^*=G$, $\widehat p_k(t)=\widehat p_k^{\ipcw}(t;G)+n^{-1}\sum_{i=1}^n f_{h_0,i}+R_n$. The weighted term is an average of independent and identically distributed terms with mean $\E[I\{C\ge D\wedge t,\,Y(t)=k\}/G\{(D\wedge t)-\}]=p_k(t)$, by the independence of $C$, and $n^{-1}\sum_{i=1}^n f_{h_0,i}$ is such an average with mean zero. Hence $\widehat p_k(t)$ is consistent for $p_k(t)$, and
\[
 n^{1/2}\{\widehat p_k(t)-p_k(t)\}=n^{-1/2}\sum_{i=1}^n\left[\frac{I\{C_i\ge D_i\wedge t,\,Y_i(t)=k\}}{G\{(D_i\wedge t)-\}}-p_k(t)+f_{h_0,i}\right]+o_p(1),
\]
so that the Aalen--Johansen is asymptotically linear with the influence function stated in the main text.

\subsection{Efficient and current-state augmentations}
Conditional on the uncensored trajectory, the mean of $I\{C\ge D\wedge t,\,Y(t)=k\}/G\{(D\wedge t)-\}$ is $I\{Y(t)=k\}$. Among subjects with $D\ge s$, conditioning instead on the history just before $s$ gives $Q_{Y(s-),k}\{s,t;\mathcal H(s-)\}$. Conditioning on the current state $j$ gives $q_{jk}(s,t)$.

For a square-integrable censoring martingale integral with predictable integrand $a(s)$ based on the history, its covariance with the IPCW contribution is
\begin{align*}
 &\E\left[\frac{I\{C\ge D\wedge t,\,Y(t)=k\}}{G\{(D\wedge t)-\}}
   \int_0^t a(s)\dd M^C(s;G)\right]\\
 &\qquad=-\int_0^t\E[I(D\ge s)Q_{Y(s-),k}\{s,t;\mathcal H(s-)\}a(s)]\dd\Lambda_C(s)
\end{align*}
because the martingale integral has mean zero and the weighted indicator vanishes when the censoring count jumps. In the compensator term, conditioning on the history removes the inverse censoring weight.

The predictable variation process $\langle M^C\rangle$ of $M^C(\cdot;G)$ satisfies
$\mathrm{d}\langle M^C\rangle(s)=I(X\ge s)\{1-\Delta\Lambda_C(s)\}\dd\Lambda_C(s)$, where $\Delta\Lambda_C(s)=\Lambda_C(s)-\Lambda_C(s-)$.
Since $\pr(C\ge s)=G(s-)$ and $1-\Delta\Lambda_C(s)=G(s)/G(s-)$, the martingale isometry shows that adding $\int_0^t Q_{Y(s-),k}\{s,t;\mathcal H(s-)\}\dd M^C(s;G)/G(s)$ cancels this covariance for every $a$. For continuous censoring this gives the full-history correction of \citet[Section~2.3]{maltzahn2026}. At a censoring atom the denominator uses the post-jump survival $G(s)$, as in \citet[Supplementary Appendix~B]{sandqvist2026}. Restricting $a(s)$ to functions of $s$ and $Y(s-)$ replaces $Q_{Y(s-),k}\{s,t;\mathcal H(s-)\}$ in the correction by its mean within the current state, $q_{jk}(s,t)$ in state $j$.

\subsection{Variance of a general augmentation}
For a fixed bounded forecast $m_k(s,t)$ based on the history just before $s$, let $\widehat p_k(t;m)$ be the sample average of
\[
 \frac{I\{C\ge D\wedge t,\,Y(t)=k\}}{G\{(D\wedge t)-\}}
 +\int_0^t\frac{m_k(s,t)}{G(s)}\dd M^C(s;G)
\]
over $n$ independent subjects. Orthogonality and the martingale isometry give
\begin{equation}\label{eq:loss}
\begin{split}
 &\var\{\widehat p_k(t;m)\}-\var\{\widehat p_k(t;Q)\}\\
 &\qquad=n^{-1}\int_0^t\E\big(I(D\ge s)[m_k(s,t)-Q_{Y(s-),k}\{s,t;\mathcal H(s-)\}]^2\big)
   \frac{\mathrm{d}\Lambda_C(s)}{G(s)} .
\end{split}
\end{equation}

For the working Aalen--Johansen prognosis in state $j$, the conditional squared error in \eqref{eq:loss} is
\[
 \var[Q_{jk}\{s,t;\mathcal H(s-)\}\mid Y(s-)=j]+
 \{P_{jk}(s,t)-q_{jk}(s,t)\}^2.
\]
The first term measures the prognostic information in the history beyond the current state. The second measures the discrepancy between the working and true state probabilities. The IPCW estimator with the true $G$ is the case $m_k(s,t)\equiv0$, for which the integrand in \eqref{eq:loss} is $\E[I(D\ge s)Q_{Y(s-),k}\{s,t;\mathcal H(s-)\}^2]$.

To obtain the geometry in Figure~2 of the main text, we set $G^*=G$, the true censoring survival function, in the exact Aalen--Johansen identity. Replacing $\widehat P$ by its limit $P$ changes the average censoring integral by $o_p(n^{-1/2})$, as shown in Appendix~C.1. Hence the influence function of the Aalen--Johansen equals the centred summand above with $m_k(s,t)=P_{Y(s-),k}(s,t)$. When $P_{jk}(s,t)=q_{jk}(s,t)$ on the support of censoring in state $j$, the difference between the weighted and Aalen--Johansen influence functions is orthogonal to the latter. The further correction for full efficiency uses the multiplier $[Q_{Y(s-),k}\{s,t;\mathcal H(s-)\}-q_{Y(s-),k}(s,t)]/G(s)$ and is orthogonal to the efficient influence function by the covariance calculation above. By Appendix~C.1, the asymptotic variance of the Aalen--Johansen exceeds that of the efficient estimator by $n$ times the right-hand side of \eqref{eq:loss} with $m_k(s,t)=P_{Y(s-),k}(s,t)$, and the decomposition of the conditional squared error given below \eqref{eq:loss} yields the variance formula of the main text.

\section{Model-free least-squares augmentation}

Throughout this section and Appendix~E, we abbreviate the score of Section~4 of the main text as $h_i(s)=h\{s;\mathcal H_i(s-)\}$ for subject $i$ and $h(s)=h\{s;\mathcal H(s-)\}$ for a generic subject. For estimation of $p_k(t)$, let the score be bounded and predictable, and let $\bar h_j(s)$ denote its sample mean among subjects at risk in state $j$ just before $s$. Let $\widehat G$ denote the pooled Kaplan--Meier estimator of the censoring survival function. With $e_i(s)=h_i(s)-\bar h_{Y_i(s-)}(s)$ the score centred within the current state, as in Section~4 of the main text, we add $\widehat\beta\widehat c_h(t)$ to the Aalen--Johansen estimate, where
\[
 \widehat c_h(t)=n^{-1}\sum_{i=1}^n\int_0^t
 \frac{h_i(s)-\bar h_{Y_i(s-)}(s)}{\widehat G(s)}\dd N_i^C(s).
\]
At each censoring time, the scores sum to zero within each state after centring. The compensators therefore cancel when $\mathrm{d}N_i^C$ is replaced by the increment of the fitted or true censoring martingale. A score constant within each state leaves the Aalen--Johansen unchanged.

The regularity conditions for Proposition~5(ii) of the main text are those of Proposition~2, a bounded predictable score, and the first-order replacement condition stated below. Under them, $n^{1/2}\{\widehat p_k(t)-p_k(t)\}$ and $n^{1/2}\{\widehat p_k^{\rm aug}(t)-p_k(t)\}$ are asymptotically normal with variances $\var(a)$ and $\var(a)-\cov(a,c)^2/\var(c)\le\var(a)$, in the notation that follows. The display in Proposition~5(ii) is this inequality between the variances of the limiting distributions. It holds for the variances of the estimators themselves, multiplied by $n$, if in addition the squares of the two normalized estimators are uniformly integrable, since their second moments then converge to those of the limits. Suppose the Aalen--Johansen estimate and the score correction are jointly asymptotically linear about $p_k(t)$ and zero, with influence functions $a$ and $c$ of finite variance. These functions account for fitting the censoring curve and within-state means. For inference on the Aalen--Johansen under a non-Markov law, see \citet{glidden2002}. Let $\mu_j(s)=\E\{h(s)\mid X\ge s,\,Y(s-)=j\}$. If replacing the fitted means and weights in the average martingale integral by their limits gives an error of $o_p(n^{-1/2})$, then
\[
 c=\int_0^t\frac{h(s)-\mu_{Y(s-)}(s)}{G(s)}\dd M^C(s;G).
\]
On a fixed grid of censoring times with positive risk sets, this follows by expanding the empirical ratios. The augmented estimate has influence function $a+\beta c$ for fixed $\beta$. If $\var(c)>0$, minimization yields
\[
 \beta^*=-\frac{\cov(a,c)}{\var(c)},\qquad
 \var\{a+\beta^*c\}=\var(a)-\frac{\cov(a,c)^2}{\var(c)} .
\]
If $\var(c)=0$, we set $\beta=0$. \citet{rubin2008} used variance minimization with the Kaplan--Meier as a benchmark, and \citet{parast2014} projected a survival estimator onto censoring martingales.

To estimate $\beta^*$ without refitting, we evaluate the two moments. Here $a$ is the centred IPCW contribution plus the integral of $P_{Y(s-),k}(s,t)/G(s)$ against $M^C(s;G)$. The latter is uncorrelated with $c$, because $P_{Y(s-),k}(s,t)$ depends on the current state alone and $h(s)-\mu_{Y(s-)}(s)$ has mean zero within each state. By the covariance calculation of Appendix~C with multiplier $\{h(s)-\mu_{Y(s-)}(s)\}/G(s)$, and since $Q_{Y(s-),k}\{s,t;\mathcal H(s-)\}$ is the conditional mean of $I\{Y(t)=k\}$ given the history,
\begin{align*}
 \cov(a,c)&=-\int_0^t\E[I(D\ge s)\{h(s)-\mu_{Y(s-)}(s)\}I\{Y(t)=k\}]\frac{\mathrm{d}\Lambda_C(s)}{G(s)},\\
 \var(c)&=\int_0^t\E[I(D\ge s)\{h(s)-\mu_{Y(s-)}(s)\}^2]\frac{\mathrm{d}\Lambda_C(s)}{G(s)}.
\end{align*}
Thus $\beta^*$ is the slope of $I\{Y(t)=k\}$ on the centred score, pooled over censoring times. In the numerator, the expectation equals that of $I(X\ge s)\{h(s)-\mu_{Y(s-)}(s)\}I\{C\ge D\wedge t,\,Y(t)=k\}/G\{(D\wedge t)-\}$; in the denominator, it equals that of $I(X\ge s)\{h(s)-\mu_{Y(s-)}(s)\}^2/G(s-)$. Replacing expectations by sample averages, $\mu_j(s)$ by $\bar h_j(s)$, $G$ by $\widehat G$ and $\mathrm{d}\Lambda_C(s)$ by the Nelson--Aalen increment $r(s)^{-1}\sum_{l=1}^n\mathrm{d}N_l^C(s)$ gives
\[
 \widehat\beta=\frac{\sum_{l=1}^n I(C_l<D_l\wedge t)\,\omega(C_l)\sum_{i=1}^n I(X_i\ge C_l)\,e_i(C_l)\,o_i(C_l)}
 {\sum_{l=1}^n I(C_l<D_l\wedge t)\,\omega(C_l)\sum_{i=1}^n I(X_i\ge C_l)\,e_i(C_l)^2},
\]
where $e_i(s)=h_i(s)-\bar h_{Y_i(s-)}(s)$ is the regressor, $o_i(s)=I\{C_i\ge D_i\wedge t,\,Y_i(t)=k\}\widehat G(s-)/\widehat G\{(D_i\wedge t)-\}$ is the response, $\omega(s)=\{r(s)\widehat G(s)\widehat G(s-)\}^{-1}$ and $r(s)=\sum_{i=1}^n I(X_i\ge s)$. This is the weighted least-squares slope described in Section~4 of the main text. If these averages converge uniformly over $s\le t$, then $\widehat\beta$ is consistent for $\beta^*$. Since $\widehat c_h(t)=O_p(n^{-1/2})$, estimating the slope then changes the estimate by $o_p(n^{-1/2})$. For a score trained on the same outcomes, its fit must also enter the joint linearization.

Suppose the score $h_i(s)$ is the true probability of occupying $k$ at $t$ given a history group that retains the current state, under the independent censoring and first-order replacement conditions of Appendix~C. Its mean within state $j$ is $q_{jk}(s,t)$, so $c$ uses the contrast $h_i(s)-q_{jk}(s,t)$. In the censoring inner product of Appendix~C, this contrast is orthogonal to any current-state prediction, and conditioning the full-history probability $Q_{Y(s-),k}\{s,t;\mathcal H(s-)\}$ on the group gives $h_i(s)$. Hence $\cov(a,c)=-\var(c)$ even if the working current-state prediction is incorrect, and for $\var(c)>0$ the optimal coefficient is one with a variance reduction of $\var(c)$ on the $n^{1/2}$ scale.

\section{Numerical experiments}

\subsection{Illness--death model}
We compare the IPCW estimator with Kaplan--Meier weights, $\widehat p_k^{\ipcw}(t;\widehat G)$, and the Aalen--Johansen estimator $\widehat p_k(t)$, with $5000$ replicates of each configuration. The first experiment uses the illness--death model of Figure~1(c) of the main text, with states $0$ (remission), $1$ (relapse) and $2$ (dead). Subjects start in state $0$, from which the $0\to1$ and $0\to2$ transitions have constant hazards $1$ and $0.3$. The sojourn time in state $1$ before death is Weibull with unit scale and shape $\kappa$, so that the process is Markov for $\kappa=1$ and semi-Markov otherwise, with a $1\to2$ hazard that decreases with the time since relapse for $\kappa<1$ and increases for $\kappa>1$. The censoring time is exponential with hazard $0.3$ or $0.8$. The target is $p_1(t)$ at $t=1.5$, which ranges from $0.26$ to $0.27$ over the four values of $\kappa$.

Table~\ref{tab:sim} shows that both $\widehat p_1^{\ipcw}(t;\widehat G)$ and $\widehat p_1(t)$ are essentially unbiased and that the variance of the former is never smaller than that of the latter. The excess reaches $17\%$ and is largest under heavy censoring and for $\kappa\le1$.

\begin{table}[!htbp]
\setlength{\belowcaptionskip}{6pt}
\caption{Illness--death model: bias and standard deviation of $\widehat p_1^{\ipcw}(t;\widehat G)$, $\widehat p_1(t)$ and $\widehat p_1^{\rm aug}(t)$ at $t=1.5$, multiplied by $10^3$}
\label{tab:sim}
\begin{tabular*}{\textwidth}{@{\extracolsep{\fill}}rrrrrrrrrrr@{}}
\toprule
 & & & \multicolumn{3}{c}{Bias} & \multicolumn{3}{c}{SD} & \multicolumn{2}{c}{Ratio}\\
 \cmidrule{4-6}\cmidrule{7-9}\cmidrule{10-11}
 $\kappa$ & Censored & $n$ & $\widehat p_1^{\ipcw}$ & $\widehat p_1$ & $\widehat p_1^{\rm aug}$ & $\widehat p_1^{\ipcw}$ & $\widehat p_1$ & $\widehat p_1^{\rm aug}$ & $\widehat p_1^{\ipcw}$ & $\widehat p_1^{\rm aug}$\\
\midrule
 0.5 & 25 & 200 & 0.5 & 0.3 & $-$0.3 & 36.9 & 35.9 & 36.0 & 1.06 & 1.00\\
 0.5 & 25 & 400 & 0.0 & 0.0 & $-$0.2 & 26.3 & 25.5 & 25.5 & 1.06 & 1.00\\
 0.5 & 50 & 200 & 0.2 & $-$0.2 & $-$3.2 & 50.8 & 47.2 & 47.3 & 1.16 & 1.01\\
 0.5 & 50 & 400 & 0.0 & 0.0 & $-$1.4 & 36.2 & 33.5 & 33.5 & 1.17 & 1.00\\
 1 & 26 & 200 & 0.5 & 0.4 & $-$0.1 & 37.3 & 36.6 & 36.6 & 1.04 & 1.00\\
 1 & 26 & 400 & $-$0.1 & 0.0 & $-$0.3 & 26.5 & 26.0 & 26.0 & 1.04 & 1.00\\
 1 & 53 & 200 & 0.5 & 0.5 & $-$2.1 & 51.9 & 49.4 & 49.4 & 1.10 & 1.00\\
 1 & 53 & 400 & 0.0 & 0.2 & $-$0.9 & 36.7 & 34.8 & 34.8 & 1.11 & 1.00\\
 5 & 29 & 200 & 0.3 & 0.4 & 0.3 & 37.5 & 37.3 & 36.6 & 1.01 & 0.97\\
 5 & 29 & 400 & 0.6 & 0.7 & 0.6 & 26.8 & 26.7 & 26.1 & 1.00 & 0.95\\
 5 & 58 & 200 & 0.0 & 0.4 & $-$0.5 & 52.4 & 51.4 & 49.6 & 1.04 & 0.93\\
 5 & 58 & 400 & 1.0 & 1.3 & 0.7 & 37.9 & 37.2 & 35.5 & 1.04 & 0.91\\
 10 & 29 & 200 & 0.3 & 0.4 & 0.3 & 38.0 & 37.9 & 37.0 & 1.01 & 0.95\\
 10 & 29 & 400 & 0.2 & 0.3 & 0.3 & 27.0 & 27.0 & 26.1 & 1.00 & 0.93\\
 10 & 59 & 200 & $-$0.3 & 0.2 & $-$0.6 & 53.0 & 52.3 & 49.5 & 1.03 & 0.90\\
 10 & 59 & 400 & 0.6 & 0.8 & 0.3 & 38.3 & 37.7 & 35.4 & 1.03 & 0.88\\
\bottomrule
\end{tabular*}
\begin{tabnote}
SD, standard deviation; Censored, percentage of subjects censored before the earlier of death and $t$; Ratio, variance divided by that of $\widehat p_1(t)$.
\end{tabnote}
\end{table}

Table~\ref{tab:sim} also reports the augmented estimator $\widehat p_1^{\rm aug}(t)$ of Section~4 of the main text, with the time spent in state $1$ as the score $h_i(s)$ and $\widehat\beta$ computed by the least-squares formula of Appendix~D. Its variance is $0.88$ to $0.97$ times that of $\widehat p_1(t)$ for $\kappa\ge5$, where the $1\to2$ hazard increases steeply with that time, and $1.00$ to $1.01$ times for $\kappa\le1$, where the dependence on duration is weaker or absent. Its bias is small relative to its standard deviation and diminishes as $n$ increases. To strengthen the dependence on duration, we raise the $0\to1$ hazard to $3$, so that subjects enter state $1$ earlier and spend longer there, and take $\kappa=5$ or $10$, for which $p_1(t)=0.16$ or $0.15$. Table~\ref{tab:aug} shows that the variance of $\widehat p_1^{\rm aug}(t)$ is then $0.84$ to $0.96$ times that of $\widehat p_1(t)$, with the largest reduction, $16\%$, at $\kappa=10$ under heavy censoring. Across Tables~\ref{tab:sim} and~\ref{tab:aug}, the reduction for $\kappa\ge5$ thus ranges from $3\%$ to $16\%$.

\begin{table}[!htbp]
\setlength{\belowcaptionskip}{6pt}
\caption{Illness--death model with $0\to1$ hazard $3$: bias and standard deviation of $\widehat p_1^{\ipcw}(t;\widehat G)$, $\widehat p_1(t)$ and $\widehat p_1^{\rm aug}(t)$ at $t=1.5$, multiplied by $10^3$}
\label{tab:aug}
\begin{tabular*}{\textwidth}{@{\extracolsep{\fill}}rrrrrrrrrrr@{}}
\toprule
 & & & \multicolumn{3}{c}{Bias} & \multicolumn{3}{c}{SD} & \multicolumn{2}{c}{Ratio}\\
 \cmidrule{4-6}\cmidrule{7-9}\cmidrule{10-11}
 $\kappa$ & Censored & $n$ & $\widehat p_1^{\ipcw}$ & $\widehat p_1$ & $\widehat p_1^{\rm aug}$ & $\widehat p_1^{\ipcw}$ & $\widehat p_1$ & $\widehat p_1^{\rm aug}$ & $\widehat p_1^{\ipcw}$ & $\widehat p_1^{\rm aug}$\\
\midrule
 5 & 27 & 200 & $-$0.4 & $-$0.4 & $-$0.6 & 31.7 & 31.6 & 30.8 & 1.00 & 0.95\\
 5 & 27 & 400 & 0.2 & 0.2 & 0.0 & 21.9 & 21.8 & 21.4 & 1.01 & 0.96\\
 5 & 56 & 200 & $-$0.2 & $-$0.6 & $-$2.4 & 44.3 & 44.0 & 41.2 & 1.02 & 0.88\\
 5 & 56 & 400 & 0.8 & 0.6 & $-$0.3 & 30.5 & 30.3 & 28.5 & 1.02 & 0.88\\
 10 & 28 & 200 & $-$0.2 & $-$0.2 & $-$0.4 & 31.1 & 31.0 & 30.0 & 1.00 & 0.93\\
 10 & 28 & 400 & 0.2 & 0.2 & 0.0 & 21.5 & 21.4 & 20.8 & 1.01 & 0.94\\
 10 & 57 & 200 & 0.5 & $-$0.1 & $-$1.9 & 43.9 & 43.5 & 39.9 & 1.02 & 0.84\\
 10 & 57 & 400 & 0.2 & 0.0 & $-$0.8 & 30.1 & 29.8 & 27.3 & 1.02 & 0.84\\
\bottomrule
\end{tabular*}
\begin{tabnote}
SD, standard deviation; Censored, percentage of subjects censored before the earlier of death and $t$; Ratio, variance divided by that of $\widehat p_1(t)$.
\end{tabnote}
\end{table}

\subsection{Progressive model}
The second experiment uses the progressive model of Figure~\ref{fig:prog}, with states $0$ (remission), $1$ (relapse), $2$ (metastasis) and $3$ (dead). Subjects start in state $0$. The $0\to1$ and $0\to3$ transitions have constant hazards $1$ and $0.1$, the $1\to2$ and $1\to3$ transitions $0.8$ and $0.15$, and the $2\to3$ transition $0.8$. If the relapse occurs before time $0.5$, all subsequent hazards are multiplied by $e^{\theta}$, so that the process is Markov only for $\theta=0$. The censoring time is exponential with hazard $0.3$ or $0.6$, and the targets are $p_k(t)$ $(k=0,1,2,3)$ at $t=2$.

\begin{figure}[!htbp]
\centering
\begin{tikzpicture}[
  x=1pt,y=1pt,>=latex,line cap=round,line join=round,
  state/.style={draw=black,rounded corners=1.6pt,line width=.4pt,
    text=black,font=\small,align=center,inner xsep=5pt,inner ysep=3pt,
    minimum height=17pt},
  transition/.style={->,line width=.6pt,shorten <=1pt,shorten >=1pt}]
  \node[state] (rem) at (0,0) {0. Remission};
  \node[state] (rel) at (115,0) {1. Relapse};
  \node[state] (met) at (230,0) {2. Metastasis};
  \node[state] (dead) at (115,-48) {3. Dead};
  \draw[transition] (rem) -- (rel);
  \draw[transition] (rel) -- (met);
  \draw[transition] (rem) -- (dead);
  \draw[transition] (rel) -- (dead);
  \draw[transition] (met) -- (dead);
\end{tikzpicture}
\caption{Progressive model.}
\label{fig:prog}
\end{figure}

The absolute bias of either estimator never exceeds $0.002$. Table~\ref{tab:prog} shows that the variance of $\widehat p_k^{\ipcw}(t;\widehat G)$ exceeds that of $\widehat p_k(t)$ by $8\%$ to $34\%$ for the transient states $k=0,1,2$ and by $2\%$ to $12\%$ for $k=3$. These results agree with Appendix~C: both estimators are augmented by a prediction that ignores the earlier history, but $\widehat p_k^{\ipcw}(t;\widehat G)$ also ignores the current state.

\begin{table}[!htbp]
\setlength{\belowcaptionskip}{6pt}
\caption{Progressive model: variance of $\widehat p_k^{\ipcw}(t;\widehat G)$ divided by that of $\widehat p_k(t)$ at $t=2$}
\label{tab:prog}
\begin{tabular*}{\textwidth}{@{\extracolsep{\fill}}rrrrrrr@{}}
\toprule
 $\theta$ & Censored & $n$ & $k=0$ & $k=1$ & $k=2$ & $k=3$\\
\midrule
 0 & 38 & 200 & 1.20 & 1.11 & 1.10 & 1.02\\
 0 & 38 & 400 & 1.18 & 1.10 & 1.10 & 1.02\\
 0 & 61 & 200 & 1.34 & 1.21 & 1.19 & 1.04\\
 0 & 61 & 400 & 1.34 & 1.20 & 1.18 & 1.04\\
 1.5 & 31 & 200 & 1.18 & 1.09 & 1.09 & 1.06\\
 1.5 & 31 & 400 & 1.16 & 1.09 & 1.09 & 1.06\\
 1.5 & 51 & 200 & 1.29 & 1.18 & 1.17 & 1.10\\
 1.5 & 51 & 400 & 1.30 & 1.18 & 1.15 & 1.12\\
 2.5 & 29 & 200 & 1.17 & 1.08 & 1.09 & 1.02\\
 2.5 & 29 & 400 & 1.15 & 1.08 & 1.10 & 1.03\\
 2.5 & 46 & 200 & 1.28 & 1.16 & 1.19 & 1.04\\
 2.5 & 46 & 400 & 1.29 & 1.16 & 1.17 & 1.06\\
\bottomrule
\end{tabular*}
\begin{tabnote}
Censored, percentage of subjects censored before the earlier of death and $t$.
\end{tabnote}
\end{table}

\end{document}